# United States v. Microsoft:
# A Failure of Antitrust in the New Economy*

By Nicholas Economides

**1.     Foreword**

The United States Department of Justice, joined by the Attorneys General of 20 States and the District of Columbia, filed on May 18, 1998 a major antitrust suit against Microsoft (DOJ Complaint 98-12320). In *United States v. Microsoft*, Microsoft ("MS") is mainly accused of:

1. Monopolization of the market for operating systems ("OSs") for Personal Computers ("PCs") through the use of anti-competitive contractual arrangements with various vendors of related goods such as with computer manufacturers and Internet Service Providers ("ISPs") and of other actions taken to preserve and enhance its monopoly, and that these acts are illegal under §2 of the Sherman Act;
2. Attempting monopolize the market for Internet browsers (but failing to succeed) which is illegal under §2 of the Sherman Act;
3. Anti-competitive bundling of the Internet Explorer ("IE"), the MS Internet browser, with the Windows operating systems, which is illegal under §1 of Sherman Act.

The district court found for the plaintiffs in almost all allegations. In particular, Judge Thomas Penfield Jackson found that:

1. The relevant antitrust market is the PC operating systems market for Intel-compatible computers, in which Microsoft has a monopoly "where it enjoys a large and stable market share."
2. Microsoft's monopoly is protected by the "applications barrier to entry," which the judge defines as the availability of an abundance of applications running Windows.
3. Microsoft used its monopoly power in the PC operating systems market to exclude rivals and harm competitors.
4. Microsoft hobbled the innovation process.
5. Microsoft's actions harmed consumers.
6. Various Microsoft contracts had anti-competitive implications, but Microsoft is *not* guilty of anti-competitive exclusive dealing contracts hindering the distribution of Netscape Navigator.

One of the key problems in the government's case was a novel theory of predation proposed (and ultimately accepted by the district court) by Professor Franklin Fisher of the Massachusetts Institute of Technology who was an expert witness for the government. According to this theory, an act is predatory if it "involves a deliberate sacrifice of profits in order to gain or protect monopoly rents as opposed to gaining of rents through superior skill, foresight, and industry."

As I discuss in detail in this paper, this theory of predation is deeply flawed for a number of reasons: (i) it can classify as predatory acts that do not involve below-cost pricing; (ii) it does not require that profits be recovered in the long run through a price increase; (iii) it is too vague and hard to implement since a judge would need to know what monopoly profits would have existed if the act that may be characterized as anti-competitive were not taken, and the extent that this particular act has affected monopoly profits; (iv) moreover, in the hypothetical long run in which the alleged anti-competitive act did not occur, profits need to be calculated when the firm has implemented other strategies not used at present because, for example, they were substitutes to the act that may be characterized as anti-competitive, or were in conflict with it; (vi) the Fisher rule requires that monopoly rents in each of the comparisons above be divided into rents that arise out of "superior skill, foresight, and industry" and those above that level, a very difficult task especially in the various hypothetical situations; and (vii) since economists have a multitude of oligopoly models and cannot agree on which is best, it may be almost impossible for the court to correctly and with high confidence make the calculations required by the Fisher rule. Use of the Fisher rule can easily characterize aggressive pro-competitive actions of firms as predatory and anti-competitive. If it is widely adopted, repeated antitrust intervention will coach aggressive competitors not be that aggressive and to instead accommodate competitors, resulting in very significant losses to consumers.

The Microsoft case is the most important antitrust case of the "new economy" thus far. Unfortunately, its legal battle was fought to a very large extent without the use of the economics tools that are at the foundation of the new economy and were key to the business success of Microsoft, as is explained in this paper. There are a number of reasons for this. First, often times legal cases are created and filed before an economist is found who will create the appropriate economic model to support the case. Second, the economic theory of networks is so inadequate and unsettled that there is no commonly accepted body of knowledge on market structure with network externalities, based on which one could evaluate deviations toward anti-competitive behavior. Third, the legal system has tremendous inertia to new ideas and models. Fourth, the legal system is ill-equipped to deal with complex technical matters. Fifth, given all these facts, lawyers on both sides find it easier to fight the issues on well-treaded ground even if the problems are really of a different nature.. I hope that in the last parts of the legal process of this case, as well as in the next new economy antitrust case, there will be a deeper understanding of the economics of networks and of the way the law should apply to network industries.

## 2.    Introduction and Brief Procedural History

This is a review of a number of issues that arise from the current major antitrust case of the United States and 19 States against Microsoft.[1] During the last few years, the

---

Editor's Note: Parts of this paper are based on Nicholas Economides, *The Microsoft Antitrust Case*, and *The Microsoft Antitrust Case: Rejoinder*, both forthcoming under Journal of Industry, Competition, and Federal Trade Commission and the Department of Justice of the United States have investigated Microsoft on various antitrust allegations. The 1991-1993 and 1993-1994 investigations by the Federal Trade Commission ("FTC") ended with no lawsuits. The 1994 investigation[2] by the United States Department of Justice ("DOJ") was terminated with a consent decree in 1995.[3] The key provisions of the 1995 consent decree were:

1.  Microsoft agreed to end "per-processor" contracts with computer manufacturers (Original Equipment Manufacturers, "OEMs") but it was allowed to use unrestricted quantity discounts.

2.  "Microsoft shall not enter into any License Agreement in which the terms of that agreement are expressly or impliedly conditioned upon the licensing of any other

---

Trade: From Theory to Policy, in August 2001. Both papers may be accessed at the personal page of the author, at http://www.stern.nyu.edu/networks/cvnoref.html.

***Nicholas Economides**: B.Sc., London School of Economics (first class honors), M.A., University of California at Berkeley, Ph.D., University of California at Berkeley. Professor Economides is currently a Professor of Economics at the Stern School of Business, New York University, New York, NY, and Visiting Professor of Economics, Stanford University, Stanford, CA, and may be reached at (212) 998-0864, or (650) 724-5270, fax (212) 995-4218, e-mail: neconomi@stern.nyu.edu, or economid@stanford.edu. For a list of publications, *see* http://www.stern.nyu.edu/networks/cvnoref.html. His web site on the Economics of Networks at http://www.stern.nyu.edu/networks/ has been ranked as one of the top 5 economics sites worldwide by The Economist magazine. His fields of specialization and research include the economics of networks, especially of telecommunications, computers and information, the economics of technical compatibility and standardization, industrial organization and the structure and organization of financial markets. Professor Economides also focuses on the application of public policy to network industries. He has published widely in the areas of networks, telecommunications, oligopoly, antitrust, product positioning and on liquidity and the organization of financial markets and exchanges. He has previously taught at Columbia University and Stanford University and has been a Visiting Scholar at the Federal Reserve Bank of New York. The author wishes to state that he is *not* a consultant of the United States Department of Justice, Microsoft, or any of the Attorneys General of the 19 States and the District of Columbia that are suing Microsoft.

[1]   Microsoft is a large diversified computer software manufacturer. Microsoft produces the Windows family of operating systems for personal computers and servers. It also produces applications software that run on the Windows family of operating systems, most notably the very successful MS-Office Suite consisting of Word (word processor), Excel (spreadsheet), PowerPoint (presentations), Outlook (e-mail and news), and Access (database). Microsoft produces software, including operating systems for PC (Windows 95, 98, NT, 2000), operating systems for local network and Internet servers (Windows NT, 2000), "back-office" products for network and Internet servers, Internet clients, Internet and network servers, desktop applications (Office, Word, Excel, Access, Outlook, PowerPoint, MS-Money, etc.), games, and programming languages (Visual Basic, Java). Microsoft also produces services, including Internet service (MSN, WebTV), Internet content (MSN), and product support, and some hardware such as branded mice, keyboards, etc. Almost all Microsoft products are complementary to a member of the Windows family of operating systems for personal computers and servers.

[2]   DOJ sued Microsoft on July 15, 1994, under 15 U.S.C. § 2 of the Sherman Act, alleging that Microsoft had entered into licensing agreements with OEMs that prevented other operating system vendors from gaining widespread distribution of their products.

[3]   The Microsoft court entered the consent decree as its Final Judgment on April 21, 1995.

Covered Product, Operating System Software product or other product (provided, however, that this provision in and of itself shall not be construed to prohibit Microsoft from developing integrated products); or the OEM not licensing, purchasing, using or distributing any non-Microsoft product."[4]

Thus, the 1995 consent decree imposes two restrictions, one horizontal, and one vertical. The horizontal restriction stops Microsoft from conditioning the payment of OEMs on the total number of personal computers they produce. However, it allows for quantity discounts to OEMs.

The vertical restriction of the 1995 consent decree prohibits product bundling created by contract, but allows Microsoft to keep expanding the number and type of functions of its products, including Windows. In short, in the 1995 consent decree contractual bundling was disallowed, but technological bundling was explicitly allowed.[5] As we will see, this issue was of crucial importance in the instant case.

During 1997, Senator Orin Hatch (R-Utah) held congressional hearings on Microsoft that featured Microsoft's CEO Bill Gates, Netscape's CEO Jim Barksdale, Sun CEO Scott McNealy, and PC manufacturer Michael Dell, among others. Senator Hatch took the position that if present antitrust law cannot deal with various anti-competitive acts attributed to Microsoft, Congress should change or enhance the antitrust laws.[6] Sun Microsystems, Oracle, IBM, Netscape, and Novell formed a loose coalition lobbying intensely for antitrust action against Microsoft.[7]

On October 20, 1997, DOJ alleged that Microsoft violated the 1995 consent decree by bundling Internet Explorer ("IE") with the Windows operating systems, and requiring computer manufacturers to distribute IE with Windows 95. DOJ petitioned the district court to find Microsoft in civil contempt. On December 11, 1997, Judge Thomas Penfield Jackson issued a preliminary injunction barring the bundling of IE with Windows.[8] On May 12, 1998, the court of appeals for the D.C. Circuit voided the 1997 preliminary injunction. On June 23, 1998, the court of appeals ruled that the 1995 consent decree did not apply to Windows 98, which was shipped with an integrated IE as part of the operating system and an IE icon on the PC desktop, arguing that "courts are ill equipped to evaluate the benefits of high-tech product design."[9]

---

[4] Final Judgment, Civil Action No. 94-1564.

[5] Microsoft has expanded over the years before and after the 1995 consent decree the functionality included in Windows, leading to the elimination of some stand-alone add-ons markets. For example, Microsoft included a disk defragmenter in Windows 1995 and the market for defragmenters promptly died. Similarly, when hard disk compression was included in Windows 1995, the market for disk compression software died. However, the market for fax software survived and expanded after the inclusion of fax capabilities in Windows 1995.

[6] *See* G. Orrin Hatch, *Antitrust in the Digital Age*, in Competition, Innovation, and the Microsoft Monopoly: Antitrust in the Digital Marketplace, Jeffrey A. Eisenach & Thomas M. Lenard (eds.), Kluwer Acad. Publishers 1999.

[7] *See, e.g.,* Gary L. Reback, *Memorandum Of Amici Curiae In Opposition To Proposed Final Judgment*, Civil Action No. 94-1564. (1994). Gary Reback represented Sun and Netscape and was instrumental in creating this loose coalition, as well as providing the main arguments that the government used in this case. *See* Joel Brinkley & Steve Lohr, *U.S. v. Microsoft* 326 (McGraw Hill 2000). The authors also mention a Netscape "white paper" written by Wilson Sonsini lawyers Gary Reback and Susan Creighton and entitled *White Paper Regarding the Recent Anticompetitive Conduct of the Microsoft Corporation*, (1996), which was never made public but was made available to the New York Times reporters.

During the week following the court of appeals stay of Judge Jackson's preliminary injunction that barred the bundling of IE with Windows because of the alleged violation of the 1995 consent decree, DOJ filed a major antitrust suit against Microsoft. In this action (DOJ Complaint 98-12320), filed on May 18, 1998, DOJ was joined by the Attorneys General of 20 States and the District of Columbia. This paper focuses on this last and continuing lawsuit against Microsoft.

Over the years, Microsoft has integrated in the Windows class of operating systems many functions and features that were originally performed by stand-alone products.[10] Moreover, the court of appeals in its June 23, 1998 decision affirmed that Microsoft's practice of bundling IE with Windows was legal under the terms of the 1995 consent decree. To overcome this interpretation of the law as far as the integration of the browser is concerned, DOJ argued that Microsoft's bundling of IE with Windows and its attempt to eliminate Netscape as a competitor in the browser market was much more than adding functionality to Windows and marginalizing a series of add-on software manufacturers. DOJ alleged (and the district court concurred) that Microsoft added browser functionality to Windows and marginalized Netscape because Netscape posed a potential competitive threat to the Windows operating system. This distinctive threat posed by Netscape was a crucial part of the DOJ allegations. DOJ alleged that applications could be written to be executed "on top" of Netscape; since Netscape could be run on a number of operating systems, DOJ alleged that Netscape could erode the market power of Windows. In DOJ's logic, Microsoft gave away IE and integrated it in Windows so that Netscape would not become a platform that would compete with Windows. Thus, DOJ alleged that Microsoft's free distribution of IE, its bundling with Windows, and all its attempts to win the browser wars were *defensive* moves by Microsoft to protect its Windows monopoly.

The Microsoft trial took place at an accelerated schedule at the U.S. District Court for the District of Columbia from October 19, 1998 to June 24, 1999. Only twelve witnesses testified from each side. Microsoft's CEO Bill Gates was not called as a witness, but his video taped deposition was extensively used during the trial. Judge Jackson pre-announced that he would announce his "findings of fact" *before* his "conclusions of law." This was widely interpreted as implying that the judge was trying to give an opportunity to the sides to reach a compromise and resolve the case through a consent decree.

---

[8] The Microsoft court also referred the issue to a special master, Professor Lawrence Lessig of Stanford.

[9] The court of appeals further noted that "the limited competence of courts to evaluate high-tech product designs and the high cost of error should make them wary of second-guessing the claimed benefits of a particular design decision," *see* 147 F.3d at 950 n.13.

[10] For example, disk compression and disk de-fragmentation were not part of Windows 3.1 and were added to Windows 95 and 98.

On November 5, 1999, Judge Jackson issued his "findings of fact," siding very strongly with the plaintiffs. In December 1999, Richard Posner, a prominent antitrust scholar and the Chief Judge of the Court of Appeals for the Seventh Circuit, agreed to serve as mediator for settlement discussions.[11] On April 1, 2000, settlement talks broke down after some States reportedly disagreed with the proposed agreement.[12] On April 3, 2000, Judge Jackson issued his "conclusions of law" finding for the plaintiffs on almost all points. In particular, Judge Jackson found Microsoft liable for monopolization and anti-competitive tying of IE with Windows but found that Microsoft's exclusive contracts did not make it liable for preventing Netscape from being distributed.

On June 7, 2000, after an extremely short hearing,[13] Judge Jackson issued his remedies decision, splitting Microsoft into two companies, and imposing severe business conduct restrictions. Microsoft appealed, and was granted a stay of all parts of the district court decisions until the appeal is heard. Although the Court of Appeals for the D.C. Circuit expressed its willingness to hear the case in plenary session, the district court agreed with the government's proposal to petition the U. S. Supreme Court to hear the case immediately, invoking a rarely used provision of antitrust law.[14] On September 26, 2000, the Court indicated that it would not hear the case before the court of appeals. Thus, the appeals process is likely to take time. It is widely believed that the federal lawsuit will be settled during the administration of President George W. Bush, who had stated during his election campaign that he will only pursue price fixing antitrust cases.[15]

The nature and scope of the allegations in this case imply that its final outcome will determine the terms of competition in the software industry and more generally in network industries. To a large extent, the application of antitrust law in network industries is hampered by the fact that rigorous economic analysis of network industries is relatively new.

---

[11] As mediator, Judge Posner was *not* acting in his judicial capacity.

[12] *See generally* N. Y. Times, Apr. 2, 2000.

[13] The hearing was on May 24, 2000. It started at 10:15 am, ended around 3:30 pm, and included a two-hour lunch break.

[14] DOJ argued that review by the U.S. Supreme Court was appropriate to expedite the final judgment because of the importance of the case for the national economy. However, it is worth noting that the plaintiffs had a real interest to avoid the Court of Appeals for the D.C. Circuit since that court ruled (June 23, 1998) on the issue of bundling Internet Explorer and Windows 95 in the earlier Microsoft case exactly in the opposite way to Judge Jackson's decision on bundling in the case being appealed.

[15] Interview with George W. Bush, Fin. Times (June 26, 2000).

## 3. The Economics of Markets With Network Effects

In assessing the Microsoft case, it is important to remember that the case focuses on markets with network effects. Network effects define crucial features of market structure that have to be taken into consideration in understanding competition and potentially anti-competitive actions in these markets.

A market exhibits network effects (or network externalities)[16] when the value to a buyer of an extra unit is higher when more units are sold, everything else being equal. In a traditional network, network externalities arise because a typical subscriber can reach more subscribers in a larger network.[17] In a virtual network,[18] network externalities arise because larger sales of component A induce larger availability of complementary components $B_1, ..., B_n$, thereby increasing the value of component A. The increased value of component A results in further positive feedback.[19] For example, the existence of an abundance of Windows-compatible applications increases the value of Windows.

There are a number of crucial features of markets with network effects that distinguish them from other markets. First, markets with strong network effects where firms can chose their own technical standards are "winner-take-most" markets. That is, in these markets, there is extreme market share and profits inequality.[20] The market share of the largest firm can easily be a multiple of the market share of the second largest, the second largest firm's market share can be a multiple of the market share of the third, and so on. This geometric sequence of market shares implies that, even for small n, the $n^{th}$ firm's market share is tiny.

For example, abundance of applications written for Windows increases the value of Windows and induces more consumers to buy Windows. This increases the incentive for independent applications writers to write applications for Windows, and this further increases sales and market share for Windows. Moreover, consumers are willing to pay more for the brand with the highest market share (since it has more associated applications), and therefore profits associated with this brand can be a large multiple of profits of other platforms. This implies a very large market share for Windows, a small market share for the Mac, a very small market share for the third competitor, and almost negligible shares for the fourth and other competitors.

---

[16] The word externality means that a good's value is not intermediated in a market. For the purposes of this paper, we will use the words "network effects" and "network externalities" interchangeably.

[17] *See* Nicholas Economides, *The Economics of Networks*, 14 Int'l J. Indus. Org. at 675-699, (visited Apr.23, 2001) at http://www.stern.nyu.edu/networks/top.html.

[18] A virtual network is a collection of compatible goods (that share a common technical platform). For example, all VHS video players make up a virtual network. Similarly, all computers running Windows 98 can be thought of as a virtual network.

[19] Despite the cycle of positive feedbacks, it is typically expected that the value of component A does not explode to infinity because the additional positive feedback is expected to decrease with increases in the size of the network.

[20] *See* Nicholas Economides & Frederick Flyer, Economides, Nicholas, *Compatibility and Market Structure for Network Goods*, Discussion Paper EC-98-02, Stern School of Business, N.Y.U., 1998, (visited Apr.23, 2001) at http://www.stern.nyu.edu/networks/98-02.pdf.

Second, due to the natural extreme inequality in market shares and profits in such markets at any point in time, there should be no presumption that there were anti-competitive actions that were responsible for the creation of the market share inequality or the very high profitability of a top firm. Great inequality in sales and profits is the natural equilibrium in markets with network externalities and incompatible technical standards. No anti-competitive acts are *necessary* to create this inequality.[21]

Third, because "winner takes most" is the natural equilibrium in these markets, attempting to superimpose a different market structure, (say one of all firms having approximately equal market shares), is futile and counterproductive. If a different market structure were imposed by a singular structural act (say a breakup of a dominant firm), the market would naturally deviate from it and instead converge to the natural inequality equilibrium. If forced equality were imposed as a permanent condition, it would create significant social inefficiency, as discussed below.

Fourth, in network markets, once few firms are in operation, the addition of new competitors, say under conditions of free entry, does not change the market structure in any significant way. The addition of a fourth competitor to a triopoly hardly changes the market shares, prices, and profits of the three top competitors.[22] This is true under conditions of free entry. Therefore, although eliminating barriers to entry can encourage competition, the resulting competition does not significantly affect market structure. In markets with strong network effects, antitrust authorities cannot significantly affect equilibrium market *structure* by eliminating barriers to entry.

---

[21] *See* Robert E. Litan, Roger G. Noll, William D. Nordhaus, & Frederic Scherer, *Remedies Brief Of Amici Curiae* (visited Apr.23, 2001), at www.aeibrookings.org/publications/related/brief.pdf. On Civil Action No. 98-1232 (TPJ) Litan *et al.* (2000) err in reasoning that Microsoft's very high profitability is a clear indication of monopolization in the antitrust sense. High profitability for the top platform is natural in this winner-take-most market).

[22] *See* Economides & Flyer, *supra* note 20. The table below, taken from this paper, shows market coverage and prices as the number of firms with incompatible platforms increases. Maximum potential sales was normalized to 1.

**Table 1: Quantities, Market Coverage, And Prices Among Incompatible Platforms**

| Number of firms I | Sales of largest firm $q_1$ | Sales of second firm $q_2$ | Sales of third firm $q_3$ | Market coverage $\Sigma^I_{j=i} q_j$ | Price of largest firm $p_1$ | Price of second firm $p_2$ | Price of third firm $p_3$ | Price of smallest firm $p_I$ |
|---|---|---|---|---|---|---|---|---|
| 1 | 0.6666 | | | 0.6666 | 0.222222 | | | 2.222e-1 |
| 2 | 0.6357 | 0.2428 | | 0.8785 | 0.172604 | 0.0294 | | 2.948e-2 |
| 3 | 0.6340 | 0.2326 | 0.0888 | 0.9555 | 0.170007 | 0.0231 | 0.0035 | 3.508e-3 |
| 4 | 0.6339 | 0.2320 | 0.0851 | 0.9837 | 0.169881 | 0.0227 | 0.0030 | 4.533e-4 |
| 5 | 0.6339 | 0.2320 | 0.0849 | 0.9940 | 0.169873 | 0.0227 | 0.0030 | 7.086e-5 |
| 6 | 0.6339 | 0.2320 | 0.0849 | 0.9999 | 0.169873 | 0.0227 | 0.0030 | 9.88e-11 |
| 7 | 0.6339 | 0.2320 | 0.0849 | 0.9999 | 0.169873 | 0.0227 | 0.0030 | 0 |

Fifth, the fact that the natural equilibrium in network industries is *winner-take-most* with very significant market inequality does not imply that competition is weak. Competition on which firm will create the top platform and reap most of the benefits is, in fact, very intense.

Sixth, there is a more fundamental concern about the application of antitrust in network industries.[23] In industries with significant network externalities, under conditions of incompatibility between competing platforms, monopoly may maximize social surplus. When strong network effects are present, a very large market share of one platform creates significant network benefits for this platform which contribute to large consumers' and producers' surpluses. It is possible to have situations where a breakup of a monopoly into two competing firms of incompatible standards *reduces* rather than increases social surplus because network externalities benefits are reduced. This is another way of saying that *de facto* standardization is valuable, even if done by a monopolist.[24]

Seventh, in network industries, the costs of entry may be higher but the rewards of success may also be higher compared to non-network industries. Thus, it is unclear if there is going to be less entry in network industries compared to traditional industries. If a requirement for entry is innovation, one can read the previous statement as saying that it is unclear if innovation would be more or less intense in network industries. The dynamics of the innovation process in the winner-take-most environment of network industries are not sufficiently understood by academic economists so that they could give

---

Note that the addition of the fourth firm onward makes practically no difference in the sales and prices of the top three firms.

[23] In the Microsoft case, both sides had the chance to address this issue, but failed to do so.

[24] Economides & Flyer, *supra* note 20, show that, in market conditions similar to the ones in the OS software market, social welfare (total social surplus) can be higher in monopoly. The table below, taken from this paper, shows profits, consumers' and total surplus in a market where firms produce incompatible products, as the number of competitors I increase.

**Table 2: Profits, Consumers' And Total Surplus Among Incompatible Platforms**

| Total number of firms I | Profits of largest firm $\Pi_1$ | Profits of second firm $\Pi_2$ | Profits of third firm $\Pi_3$ | Total industry profits $\Sigma_{i=1}^{I} \Pi_i$ | Consumers' surplus CS | Total surplus TS |
|---|---|---|---|---|---|---|
| 1 | 0.1481 | | | 0.1481 | 0.148197 | 0.29629651 |
| 2 | 0.1097 | 7.159e-3 | | 0.1168 | 0.173219 | 0.29001881 |
| 3 | 0.1077 | 5.377e-3 | 3.508e-4 | 0.1135 | 0.175288 | 0.28878819 |

credible advice on this issue to antitrust authorities. However, in the last two decades we have observed very intense competition in innovative activities in network industries financed by capital markets.

Eighth, the existence of an installed base of consumers favors an incumbent. However, competitors with significant product advantages or a better pricing strategy can overcome the advantage of an installed base.[25] Network effects intensify competition, and an entrant with a significantly better product can unseat the incumbent. In network industries, we often observe *Schumpeterian* races for market dominance. This is a consequence of the winner-take-most natural equilibrium combined with the high intensity of competition that network externalities imply.

**4.     The Allegations**

On May 18, 1998, the United States Department of Justice, and the Attorneys General of 20 States[26] and the District of Columbia sued Microsoft. The main allegations were:

1. Microsoft illegally monopolized the market for operating systems ("OSs") for personal computers ("PCs") under § 2 of the Sherman Antitrust Act;

2. Microsoft had anti-competitive contractual arrangements with various vendors of related goods, such as with computer manufacturers ("OEMs") and Internet Service Providers ("ISPs"), and had taken other actions to preserve and enhance its monopoly; that these contractual arrangements and other actions were illegal under §2 of the Sherman Antitrust Act;

3. Microsoft illegally attempted to monopolize the market for Internet browsers (but failed to succeed), an act that is illegal under § 2 of the Sherman Antitrust Act;

4. Microsoft bundled anti-competitively its Internet browser, IE, the Microsoft Internet browser, with its Windows operating systems; that this is illegal under §1 of the Sherman Antitrust Act.

---

[25]     A clear example of this is the win of VHS over Beta in the United States consumer video recorders market. Beta was first to market and had a significant installed base in the five years of the coexistence of the two competing standards. However, because VHS (i) introduced earlier a recording tape of longer duration; (ii) used wide and inexpensive licensing of its technology; and (iii) its licensees had a much wider distribution system, VHS emerged as the winner, and Sony stopped selling Beta recorders to the US consumer market.

[26]     The plaintiff States were California, Connecticut, Florida, Illinois, Iowa, Kansas, Kentucky, Louisiana, Maryland, Massachusetts, Michigan, Minnesota, New Mexico, New York, North Carolina, Ohio, South Carolina, Utah, West Virginia and Wisconsin. Originally South Carolina was part of the lawsuit but dropped out of the case soon after the acquisition of Netscape by America Online ("AOL") and Sun Microsystems ("Sun") stating that this acquisition has eliminated any anti-competitive concerns.

### 5. Microsoft's Defense

Microsoft's defense was as follows. First, Microsoft argued that the law was on its side since the court of appeals had ruled on June 23, 1998 that Microsoft can legally add new features and functions to Windows. Therefore Microsoft argued that it was legal to add IE's functionality to Windows, and it had done nothing wrong by integrating IE in Windows.

Second, Microsoft argued that it was just competing hard against Netscape, that such competition was welfare-enhancing, and that it did not commit any anti-competitive acts.

Third, Microsoft argued that it did not have monopoly power in the operating systems market.

Fourth, Microsoft argued that competition in the software sector was intense and that its leadership position could be replaced at any time by a new competitor or entrant. Although this seems to have been a deeply held belief of Microsoft's management (as revealed by the internal e-mails), its economics expert witness failed to convincingly articulate this *Schumpeterian* view in the context of an antitrust defense.

Fifth, Microsoft argued that it is a leader in software innovation and that it has enhanced rather than hobbled the innovation process.

Sixth, Microsoft argued that consumers have benefited from its actions rather than been harmed by them. Microsoft claimed direct consumer benefits from its low pricing of the operating system, the zero pricing of its Internet browser, and from its enhancement and acceleration of the innovation process. Microsoft also argued (rather ineffectively) that consumers benefit from the *de facto* standardization that its large market share brought to the operating systems market.

### 6. Antitrust Law

The analysis of DOJ's allegations and Microsoft's defenses requires an examination of antitrust law as it currently applies in the United States, the facts of the case, and a synthesis that will define the extent of the violations of the law. We start with the examination of antitrust law as it applies to this case.

We review the antitrust law in the areas that apply to the Microsoft case. Section 2 of the Sherman Antitrust Act states:

> "Every person who shall monopolize, or attempt to monopolize, or combine or conspire with any other person or persons, to monopolize any part of the trade or commerce among the several States, or with foreign nations, shall be deemed guilty of a felony, and, on conviction thereof, shall be punished by fine not exceeding $10,000,000 if a corporation, or, if any other person, $350,000, or by imprisonment not exceeding three years, or by both said punishments, in the discretion of the court."

The U.S. antitrust law, as presently interpreted, implies that "monopolization" under § 2 of the Sherman Act is illegal if the offender took anti-competitive actions to acquire, preserve, or enhance its monopoly. For "monopolization," plaintiffs have to prove that the defendant:

1. Possessed market power; *and*
2. willfully acquired or maintained this monopoly power as distinguished from acquisition through a superior product, business acumen, or historical accident.

Therefore, contrary to popular belief, for monopolization to be illegal under U.S. antitrust law, it is not sufficient for a company to "monopolize" a market in the sense of possessing a very large market share, even a market share of 100%.

U.S. antitrust law, as presently interpreted, implies that "attempting to monopolize" is illegal under § 2 of the Sherman Antitrust Act if the specific actions taken have anti-competitive consequences. For example, bundling, and, more generally, price discrimination could be illegal if they have anti-competitive consequences. Similarly, exclusionary contracts (which restrict distribution or production) could be illegal if they have anti-competitive effects.

To prove "attempting to monopolize" (under Sherman Act § 2), plaintiffs have to prove that the defendant:

1. Engaged in predatory or anti-competitive conduct
2. with specific intent to monopolize
3. and that there was a "dangerous probability" that the defendant would succeed in achieving monopoly power.

Section 1 of the Sherman Antitrust Act states:

"Every contract, combination in the form of trust or otherwise, or conspiracy, in restraint of trade or commerce among the several States, or with foreign nations, is hereby declared to be illegal. Every person who shall make any contract or engage in any combination or conspiracy hereby declared to be illegal shall be deemed guilty of a felony, and, on conviction thereof, shall be punished by fine not exceeding $10,000,000 if a corporation, or, if any other person, $350,000, or by imprisonment not exceeding three years, or by both said punishments, in the discretion of the court."

Thus, unreasonable "restraint of competition" is illegal under §1 of the Sherman Act; this may include tying of products or other exclusive arrangements.

It has generally been accepted in the application of antitrust law in the United States that anticompetitive acts harm consumers. Consumer harm or attempted harm is necessary for an antitrust violation. The affected consumer group may be present or future. Consumers may lose directly from high prices or indirectly through a limitation of choices of variety and quality or by a retardation of the innovation process. But without consumer victims or prospective likely victims, present or future, it is extremely hard to prove that an antitrust violation exists.

## 7. The Judge's "Findings Of Fact" And "Conclusions Of Law"

The U.S. District Court's "findings of fact" (November 1999) and "conclusions of law" (April 2000) find for the plaintiffs (U.S. Department of Justice and 19 States) in almost all the allegations against Microsoft. In particular, Judge Penfield Jackson found:

1. The relevant antitrust market is the PC operating systems market for Intel-compatible computers.
2. Microsoft has a monopoly in this market "where it enjoys a large and stable market share."
3. Microsoft's monopoly is protected by the "applications barrier to entry," which the judge defines as the availability of an abundance of applications running Windows.
4. Microsoft used its monopoly power in the PC operating systems market to exclude rivals and harm competitors.
5. Microsoft hobbled the innovation process.
6. Microsoft's actions harmed consumers.
7. Various Microsoft contracts had anti-competitive implications, but Microsoft is *not* guilty of anti-competitive exclusive dealing contracts hindering the distribution of Netscape Navigator.

## 8. On Monopolization Of The Operating Systems Market
### a. Market Definition

The definition of a market for antitrust purposes is crucial for the determination of liability. Traditionally, an antitrust market definition is based on substitution considerations. To define the market one has to examine the product. Windows was sold to OEMs and users as an operating system on which they could run software applications. Windows also contained a variety of functions that were provided free of charge to independent software vendors ("ISVs") so that they would find it easier, cheaper, and faster to create applications for Windows. Microsoft sold to ISVs various tools so that they would utilize these functions of Windows, and provided a variety of free services to ISVs. Clearly ISVs were subsidized in kind so that they would write applications for Windows. DOJ and the district court decided to ignore the market relationship between ISVs and Microsoft and to focus exclusively on the sale of Windows to OEMs.

DOJ proposed and the district court judge agreed that the relevant antitrust market was the market for operating systems for personal computers that are based on an Intel-compatible Central Processing Unit ("CPU"). In this respect, the judge took a rather narrow and static view of the market. It is clear that consumers demand applications services from personal computers and would be willing to switch their purchases to non-Intel-compatible PCs if these computers were offered with similar applications as Intel-based PCs and at attractive prices. Thus, non-Intel-compatible computers, operating systems and software are substitutes for Intel-compatible computers, operating systems and software. If one takes a static view and considers the software applications written for each computer and operating system as given and unchanging (as the judge did), then there is no close substitute for Windows that can run as many applications as Windows does. In this static view, all substitutes for Windows are much less valuable because they can run so few applications compared to Windows. In a more dynamic view, if new software applications can be written for an operating system (or old ones originally written for other operating systems "ported to" it), an operating system can easily have close substitutes.

The dynamic view of market definition invites an examination of the incentives that ISVs have to create such applications. Their incentives are not just defined by the price of the operating system to OEMs and final consumers, and by the number of units that the operating system is expected to sell over the lifetime of their applications. ISVs use functions that are embedded in an operating system or platform to perform certain operations that are necessary for their applications. The more such functions are available in an operating system, and the easier they are to access, the more likely it is for an ISV to write an application for the OS. Thus, the availability, abundance, and ease of use of embedded functions in the operating system that can be utilized by an ISV needs to be calibrated in order to define the extent of the market in the dynamic view of market definition.

Early in the trial, Microsoft argued that the concept of market definition for antitrust purposes was inappropriate for the dynamic setting of software markets. This position was not very well articulated, and hardly appropriate for a district court judge who was very unlikely to subscribe to a major revision to traditional application of antitrust law. Later in the trial, Microsoft argued through its economic witness that *platforms* rather than operating systems were the appropriate domain for antitrust analysis, stressing the fact that functions embedded in platforms are used by ISVs to cut their costs of writing applications. However, a full and coherent view of a *dynamic* market definition was never presented.

### b. Barriers To Entry

A crucial tenet of DOJ's and the district court's argument on monopolization was that there are significant barriers to entry in the market for operating systems for personal computers. Barriers to entry are asymmetric costs that entrants incur but are not incurred by incumbents. Software has a very large fixed cost and a very low (almost zero) marginal cost. The fixed cost required to create an operating system could create a barrier to entry to the extent that it is sunk by the incumbent(s), while it has to be paid by an entrant in the current period. However, this barrier to entry is not very significant, given the size of the market.

The main barrier to entry could be created by the abundance of applications that run on Windows. It is estimated that Windows runs 70,000 applications, while the Macintosh and the Linux operating systems run far fewer applications. Currently, there are also applications that run on servers connected with the World Wide Web and are essentially independent of the operating system (Windows, Mac, Linux, etc.) of the client. The number of applications that can be run on an operating system can be thought of as a quality index. What determines the existence and the extent of the "applications barrier to entry" is the extent that costs incurred today by entrants are already sunk by the incumbent. The fact that an industry has a high fixed cost or requires investment to induce others to create complementary goods, *by itself* is not enough to conclude that there are barriers to entry. To judge whether barriers to entry exist, one has to examine if, in the relevant time frame, the entrants are facing *asymmetrically* higher costs than the incumbent(s).

It is clear that, early in its history, Microsoft understood the importance of an abundance of applications for its operating systems. To achieve this, Microsoft provided various inducements to independent applications developers to write applications for its operating systems. A key inducement for independent applications developers has been the availability of functions in Windows that are useful for applications developers. For example, Windows and other operating systems include programming code that makes it easy for an application to print to any of a large number of printers.[27] Since this function is available in the operating system, applications developers do not have to write their own programming code to accomplish this function. It follows that applications are now cheaper and quicker to write.

Microsoft spends considerable resources to attract independent software developers to write for its operating system and to include in the operating system a variety of functions that they find useful. It is clear that the abundance of applications for Windows makes Windows more valuable, and therefore increases Microsoft's potential profits. This is clearly a pro-competitive reason for Microsoft to invest in resources to attract independent software developers. Given the same number of competitors in the operating systems market, a firm enhances its competitive position by investing resources in order to attract applications to its operating system. At the same time, such an increase in the number of applications available for an operating system makes it very difficult for new entrants to profitably enter and survive. Thus, increasing the number of applications available for an operating system, although it may have a purely competitive justification, both increases the cost of entry in the operating systems market, and creates the potential for anti-competitive exploitation.

---

[27] This function was not available in DOS. Thus, each application written for DOS had to include large numbers printer drivers that made sure that the application would be able to print on users' printers.

On the other hand, to the extent that the incumbent is *currently* spending resources to attract applications to its operating system, the *difference* between the costs faced by the incumbent and a potential entrant is *reduced,* and therefore barriers to entry are *reduced*. Thus, the existence, extent, and importance of an "applications barrier to entry" depends on an empirical examination of the extent to which Microsoft spends resources in the current time frame to attract applications to Windows, and the extent to which Microsoft has already sunk such costs in past time periods. Such an examination was not performed by the district court.

### c. Existence Of Market Power And Pricing Of The Operating System

The judge ruled that Microsoft has monopoly power in the OS market for Intel-compatible PCs. In antitrust, it is generally understood that a firm has monopoly power when it has the sustained ability to control price or exclude competitors. The existence of significant barriers to entry and the very high market share of Microsoft in the operating systems market gave indications that Microsoft had monopoly power. But there was also a very strong indication to the contrary. Microsoft priced its operating system to Original Equipment Manufacturers ("OEMs") at an average price of $40-60, a ridiculously low price compared to the static monopoly price.[28] Microsoft's economic witness showed that the static monopoly price was about $1,800, a large multiple of Microsoft's actual price.[29] At first blush, it seems that Microsoft could not possibly have monopoly power in OSs when its OS price is about 3% of the monopoly price.

---

[28]    It is very likely that the marginal price for the last unit sold to the same OEM was extremely low since the 1995 consent decree allowed Microsoft to have quantity discounts but barred it from zero marginal cost pricing.

[29]    The derivation of the monopoly price for Windows follows. Let $p_H$ be the price of the PC hardware (everything except Windows) and let $p_W$ be the price of Windows. Assume that Windows is installed on all PCs, i.e., that Microsoft has 100% market share. Since hardware and software are combined in a ratio of 1:1, the demand of a PC with Windows is $D(p_H + p_W)$. Profits of Microsoft from Windows sales are:

$$\Pi_W = p_W D(p_H + p_W) - F_W$$

where $F_W$ is the fixed cost of developing Windows, and the marginal cost is negligible. Maximizing $\Pi_W$ implies marginal revenue equals marginal cost, i.e.,

$$D(p_H + p_W) + p_W \, dD/dp_W = 0 \iff 1 + [p_W/(p_H + p_W)][(p_H + p_W)/D][dD/dp_W] = 0 \iff p_W/(p_H + p_W) = 1/|\varepsilon|,$$

or equivalently, the monopoly price of Windows is

$$p_W = p_H/(|\varepsilon| - 1),$$

where $|\varepsilon| = -[(p_H + p_W)/D]/[dD/dp_W]$ is the market elasticity of demand for PCs with Windows. If one assumes that the average price of PC hardware is $1,800 and the elasticity is $|\varepsilon| = 2$, the monopoly price of

Understanding and explaining the very low price of Windows is important for understanding what Microsoft's competitive position was and how Microsoft thought of it. Plaintiffs failed to explain Microsoft's pricing by either assuming it away or giving spurious explanations. First, the government claimed that pricing significantly above marginal cost was evidence of monopoly power, and that the discrepancy between actual and theoretical monopoly price did not matter. But, any software is priced significantly above marginal cost since marginal cost is zero, and the government has not yet sued other software manufacturers on these grounds. Second, the government's economic witnesses claimed that the static monopoly model of the plaintiffs did not apply, but offered no alternative model that could explain the difference between the actual price and the monopoly price of the static model. Third, the government claimed that a large market elasticity and revenues from complementary goods were sufficient to justify that the actual price was equal to the monopoly price. But the required elasticities are very large and inconsistent with the price to cost margins of *hardware* manufacturers.[30]

Why was the price of Windows low? For the early periods of each operating system, one expects that the existence of network effects would prompt Microsoft to charge a low price so that each platform becomes accepted by independent software developers as well as users and the bandwagon gets rolling. This network-effects-based theory does not explain why Microsoft did not increase the prices of each generation of operating systems as each one matured. It also does not explain why Microsoft did not increase significantly its price of Windows as it doubled its market share in the past four years.

A variation of the network effects-induced low price theory gives a predatory flavor to Microsoft's strategy. In this view, Microsoft priced low to hook consumers and generate network effects, while it planned to increase the price "in the future." This theory is particularly implausible because Microsoft dominates the PC desktop market and has doubled its market share in recent years without increasing the price for Windows while expanding its functionality. How long will Microsoft wait until it increases the price of Windows? Some even claim that in a network industry, a firm can practice predation without *ever* increasing the price, but just benefiting in the future from the network effects. But such a strategy is indistinguishable from a truly competitive strategy and cannot be considered predatory.

A number of other theories have been proposed to explain Microsoft's pricing. Some claim that the availability of an existing installed base of Windows constrained Microsoft's pricing. That could have been true if in fact Windows could easily and

---

Windows is $p_W = \$1,800$. Even if one assumes a much higher elasticity, $|\varepsilon| = 3$, and a much lower average price of PC hardware at $1,200, the monopoly price is $600, which is ten to twelve times the price charged by Microsoft to OEMs.

[30]    In the Windows monopoly price formula $p_W = p_H/(|\varepsilon| - 1)$, if the PC hardware price of $1,800 one requires an elasticity of $|\varepsilon| = 31$ to get a Windows monopoly price of $60. A PC hardware price of $1,200 one requires an elasticity of $|\varepsilon| = 21$ to get a Windows monopoly price of $60. It is extremely unlikely that that the market for PCs exhibits such large market elasticities.

legally be uninstalled by users. But Microsoft's licensing requirements and the sheer complexity of the uninstallation operation make it almost impossible for a user to uninstall a Windows operating system that was installed by an OEM and move it to a different (presumably new) computer. So the Windows installed base does not constrain Windows pricing.

Others claim that the fact that software is durable constrains Windows pricing. It is true that once Windows runs on an overwhelming market share of PCs, the substitute for a new Windows computer is an old Windows computer. The fact that computers and software are both durable makes this true. But very rapid technological change has prompted consumers to buy new computers much faster than traditional obsolescence rates would imply. Even doubling the price of Windows to OEMs would not have implied a significant change to the final price of the combined computer and operating system. So, durability of software and durability of computers, although a factor, is unlikely to explain the huge difference between the actual price and the static monopoly price.

Still others claim that the low price is implied by the very low cost of pirating software. If that were the case, it would have prompted Microsoft to cut its much higher prices of MS-Office and other software, since pirating takes the same effort and has the same costs for pirates irrespective of the type of software being pirated. Moreover, the control that Microsoft can exert on piracy of the operating system is much greater than on piracy of applications. Therefore, although piracy could have been more of a problem if the OS software was much more expensive, it is unlikely that the price of Windows is low because of piracy considerations.

Finally some claim that Microsoft kept the price of Windows low because that allowed Microsoft to charge more for complementary goods that it produces, such as the Microsoft Office suite. There are three reasons that make this argument unlikely to be correct. First, Microsoft also produces its most popular products, including the Microsoft Office suite, for the Mac. If Microsoft kept the price of Windows low so that is sold MS-Office for Windows at a high price, then the price of MS-Office for Macs should have been lower than the price of MS-Office for Windows, which is factually incorrect. Second, Windows has the ability to collect surplus from the whole assortment of applications that run on top of it. Keeping Windows' price artificially low would subsidize not only MS-Office, but also the whole array of tens of thousands of Windows applications that are *not* produced by Microsoft. Therefore, even if Microsoft had a monopoly in the Office market, keeping the price of Windows low is definitely *not* the optimal way to collect surplus. Third, even receiving large per unit revenues from complementary goods cannot by itself explain the vast difference between the actual and the monopoly price of Windows.[31]

---

[31] Suppose Microsoft receives net revenue $R_C$ sold from goods that are complementary to Windows for every unit of Windows sold. Then its profits are

$$\Pi_W = (p_W + R_C)D(p_H + p_W) - F_W .$$

Then the monopoly price of Windows is

Microsoft claimed that its low pricing was due to actual and potential competition. Microsoft's Internal e-mails point to a real fear that the company would be overtaken by the next innovator. Even if objectively it is difficult to see the big threat from potential competitors, it is clear that Microsoft's executives constantly felt the fear of potential competition. On the pricing of Windows, I am inclined to believe Microsoft's view: Microsoft priced low because of the threat of competition. This means that Microsoft believed that it could not price higher if it were to maintain its market position. In essence, Microsoft pricing reveals that to a large extent its executives believed that market conditions (most importantly potential competition) constrained higher prices.

### d. Exercise Of Market Power

The government claimed that Microsoft's actions to exclude and marginalize Netscape were sufficient to show that Microsoft possessed and exercised its monopoly power. Essentially the government claimed that even if Microsoft did not exercise monopoly power in pricing, it was sufficient to show that it exercised it in exclusionary actions. But, most economists would agree that it is much more profitable to exercise market power by increasing price than by raising the costs of rivals.[32] If Microsoft consistently sells at low prices, it loses a large amount of potential profits. It is very hard to make a convincing argument that it is worth sacrificing these profits just to exclude a future potential competitor. After all, even if the potential competitor is successful in entering, it could only reduce Microsoft's future profits, which are worth less today.

The government's and the judge's theory was that Microsoft exercised its monopoly power by attempting to marginalize Netscape. The government and the judge agree that Microsoft attempted to marginalize Netscape's browser because Microsoft feared that Netscape would become a rival platform to Windows. According to this point of view, once Netscape became such a platform, applications would be written to run on Netscape Navigator. Moreover, the bigger Netscape's market share, the more likely it is that ISVs would write applications for the Netscape platform; it follows that, in DOJ's view, competition in OSs is maximized when Netscape has a monopoly in browsers. Since Netscape Navigator could run on many operating systems (not just Windows), therefore, from this perspective, Netscape created a threat to Windows. In this view, all actions to aggressively compete with Netscape were just attempts by Microsoft to defend its monopoly in PC operating systems.

---

$$p_W = p_H/(|\varepsilon| - 1) - R_C .$$

Even if Microsoft makes a net revenue $R_C = \$200$ from complementary goods from every unit of Windows sold, when $p_H = \$1,800$ and $|\varepsilon| = 2$, the monopoly price of Windows is $p_W = \$1,600$, and when $p_H = \$1,200$ and $|\varepsilon| = 3$, the monopoly price of Windows is $p_W = \$400$, vastly above the actual price.

[32] The exception is regulated industries where a firm is not free to increase price.

This is an interesting theoretical argument, but it has both theoretical and empirical deficiencies. First, in winner-take-most competition, DOJ's analysis fails to distinguish between profit sacrifices to keep out potential rivals and competition where a company keeps its prices low in order to win consumer business from actual and potential rivals. Second, although at some point in time Marc Andreesen, Netscape's CTO, claimed that Netscape would create the capabilities of running applications, there was never a consistent effort by Netscape to create a comprehensive set of Applications Product Interfaces ("APIs") that would support typical applications that now run under Windows. Moreover, Netscape hardly put sufficient effort and resources in reaching independent applications developers to write for such a platform. So, although potentially Netscape could have created such a platform, it was very far away from that goal, and the probability of succeeding against the array of Windows applications was low.

Third, Netscape, even if successful, is likely to have been limited to Internet-based applications. The performance of such applications is dependent on the Internet working efficiently and at top speed. Actual performance would also potentially be hampered by excessive loads on the CPUs of servers where the applications would actually run. Essentially the model of Netscape as competitor of operating systems required a number of contingencies to be met, and therefore faced a very significant uncertainty. If instead applications were to run on the PC itself (on top of Netscape) they would need to interface to some extent with the underlying operating system. If this system were Windows, then it would be unlikely that it would accommodate perfectly its direct competitor.

Fourth, Microsoft's actions clearly could have had competitive justifications. Microsoft's management saw that the Internet was taking off and wanted to have a dominant position in it.

DOJ and the district court judge also agreed that Microsoft sabotaged Sun's Java operating language so that Java would not become a "universal language" in the sense that applications written for it could be run in any operating system. Over the years, the computer industry has gone through a number of such pipe dream "universal languages." They have all failed for a very simple reason: a "universal language" cannot be too specialized—after all it has to be recognizable by diverse operating systems. But efficiency and speed in program execution requires specialization. Universal languages eventually fail because they get more and more efficient for particular operating systems and thereby their universality fails.[33]

Microsoft created an implementation of Java for Windows that was more efficient than the original Java by Sun. Microsoft's implementation (that is, Microsoft's Java virtual machine) allowed for all programs written for the original ("pure") Java to be run on it. Thus, it preserved *backward compatibility* with the original Java that ran on all operating systems. Because of that, Microsoft's actions were not anti-competitive. Microsoft claims that Sun's non-OS-specific Java language was inefficient and slow, and that Microsoft improved it. This is a reasonable pro-competitive justification for Microsoft's actions. The fact that Microsoft's Java implementation also runs programs that do not run in other Java implementations is not cause for anti-competitive concern, especially since Microsoft has published the function calls its Java makes to Windows. Competing operating systems could implement these calls.[34]

---

[33] The fate of human languages is similar. As each language becomes more efficient for a certain group it gets more specialized and cannot easily be understood by the rest of the world.

### e. A Novel Theory of Predation

A key economic witness of DOJ, Professor Frank Fisher of MIT, proposed in pre-trial filings and supported during the trial a novel theory of predation. The traditional theory of predation requires that a product be sold below incremental (or "avoidable") cost for a period of time until competitors are driven out of business; then the monopolist increases prices and reaps its monopoly profit.[35] Instead Professor Fisher calls an action predatory if it would not have been profitable (and therefore not undertaken) in the long run unless the predator was taking into consideration profits arising out of the negative impact of the practice on competition.

Specifically Professor Fisher's standard (from pre-filed testimony in the Microsoft case and from the cross-examination of Professor Fisher) is that an action is anti-competitive when such an action "would not be profit maximizing except for the monopoly rents earned or protected by the action." In a comment to my earlier criticism of this predation theory, Professor Fisher[36] augmented his definition/rule ("new Fisher rule" thereafter) of an anti-competitive act as one that "involves a deliberate sacrifice of profits in order to gain or protect monopoly rents as opposed to gaining of rents through superior skill, foresight, and industry," essentially adding the part "as opposed to the gaining of rents through superior skill, foresight, and industry."

According to this definition of predation, it can occur even when the predator sets a price above incremental cost, or even above average unit cost.[37] All that is required for predation, according to Professor Fisher, is for a company to take some (presumably very aggressive) action that would not be rewarding unless competitors are hurt. If this definition is to be adopted, a large array of actions that are normal business practices and which benefit consumers are going to be characterized as predatory and therefore illegal.

---

[34]   This discussion is limited to antitrust considerations and does not discuss the *contractual* dispute between Microsoft and Sun.

[35]   *See, e.g.,* Phillip Areeda & Louis Kaplow, *Antitrust Analysis* 512, 514 (Aspen L.& Bus. 1997).

[36]   *See* Fisher *Innovative Industries and Antitrust: Implications of the Microsoft Case*, J. Indus., Competition and Trade: From Theory to Policy (2001).

[37]   Keep in mind that software has a very low (almost zero) incremental cost and a decreasing average (unit) cost that is always higher than incremental cost.

In antitrust economics, it is widely believed that rules should be fashioned to maximize social surplus. The Fisher rule fails to do so, as I show below. It creates an artificial cushion of profits. If an aggressive competitor dares to take away some of these profits, he is named a predator. Moreover, since the profit cushion is created without regard to cost, it can be at any *status quo* level. Instead of promoting competition and efficiency, the Fisher predation theory could protect slack and monopoly rents. As such, it would appear that the judge erred to the extent that he based his *Microsoft* ruling on this incorrect theory.

To better understand Professor Fisher's definition of predation, consider the following example. Suppose that in a two-firm industry, firm 1 has constant average unit cost equal of $10, while firm 2 has constant average unit cost of $15. An "accommodation strategy" of firm 1 would be to charge a price above $15 (say $16), so that the market is shared between firms 1 and 2. Suppose that firm 1 plays an "aggressive strategy" and charges a unit price of $14. Such an action results in a profit for firm 1 of $4 for every unit sold in the market. These profits are likely to be higher than the profits firm 1 would have received in the accommodation strategy of $16. The accommodation strategy gives a profit of $6 per unit sold by firm 1 and, since firm 1 sells only half the output, its profit is $3 for every unit sold in the market. If the market is not extremely elastic, firm 1 makes more profits if it uses the aggressive strategy, which drives firm 2 out of business. Applying Professor Fisher's predation definition, firm 1 is a predator because its aggressive action increased its monopoly rent. Moreover, in comparison with the "accommodation strategy," the aggressive strategy would have not been taken if it were not trying to increase its monopoly rent.

If the "aggressive strategy" is considered predatory and therefore illegal according to Professor Fisher's criterion and firm 1 uses the "accommodation strategy" instead, total and consumers' surplus will decrease. Consumers will face a higher price of $16 rather than $14 and will therefore be worse off. Moreover, firm 2 will be producing half the units sold at a cost that is higher by $5 than the most efficient production cost, thereby reducing total surplus. It is evident that application of Professor Fisher's predation criterion can result in lower consumers' and total market benefits (total surplus). Moreover, the application of this criterion shields the less efficient firm from competition and guarantees its rents.

Additionally, the Fisher rule is vague and hard to implement. To implement this rule, the antitrust authority, a judge, or a jury need to know what monopoly profits would have existed if the act that may be characterized as anti-competitive were not taken, and the extent that this particular act has affected monopoly profits. This is a task of enormous difficulty and uncertainty. The judge will have to be able to calculate the profits that would have been realized in a hypothetical situation in which the particular act (that may be possibly characterized as anti-competitive) were not taken, compare them with the present situation where the act was taken, and conclude that profits were lost as a consequence of taking the particular act. This requires a calculation of the firm's profits in the long run when the act is taken and profits in the long run when the action is not taken. And, a proper comparison would require a calculation of the firm's profits in the long run when, having not taken the act that may be characterized as anti-competitive, the firm implemented instead other strategies. That is, if one makes long run maximum profits comparisons, and truly wants to implement Fisher's rule, one needs to calculate the maximized profits in the absence of the act that may be characterized as anti-competitive. This requires calculating profits where the firm has implemented other strategies not used at present.[38]

This task is extremely difficult because it requires that the judge and jury understand the long run optimization strategy of the firm. Economists have a multitude of oligopoly models and cannot agree on which is best. It may be almost impossible for the court to correctly and with high confidence make the calculations required by the Fisher rules. In the Microsoft case, for example, both sides failed to adequately explain the long run profit maximization model that Microsoft actually used to price Windows over a long period of time, before as well as after the alleged anti-competitive actions. One wonders how the courts would possibly be able to make the additional hypothetical calculations required for the implementation of the Fisher rule. What oligopoly model would the court adopt? What other strategies would the court expect to have been implemented in the absence of the act that may be characterized as anti-competitive? What equilibrium paths would the court expect to be followed in the long run? These are very hard questions that are unlikely to be answered in a satisfactory way given our present knowledge of economics. In practice, the Fisher rule would constrain dominant firms from aggressive pricing, investment, and innovation behavior that would be seen, at least by some economists, as fitting exactly the requirements of anti-competitive behavior of the Fisher rule. If such aggressive behavior is prevented or punished, the ultimate losers are current and future consumers.

Moreover, the additional caveat of the new Fisher rule means that the present versus hypothetical analysis described above has to be done with reference to monopoly rents over and above those that arise out of "superior skill, foresight, and industry." So, two types of monopoly rents have to be calculated in each case of the hypothetical equilibrium (where the action is not taken but possibly other hypothetical actions are taken) and of the actual equilibrium (where the action was taken). The judge and jury now have the additional burden to ascertain the extent that present as well as hypothetical rents are monopoly rents not arising out of "superior skill, foresight, and industry." This makes their task even more difficult.

Based on these considerations, the Areeda predation rule is vastly superior because it does not require the court to make the very difficult profit maximization calculations described above, some of which are by their nature speculative because they require an analysis of firm behavior in hypothetical situations. Instead, the Areeda rule is based on a comparison of incremental cost and price, which are much easier to determine than what is required under the Fisher rule.

---

[38] Such strategies might not have been implemented because, for example, they were substitutes to the act that may be characterized as anti-competitive, or were in conflict with it.

There is a second important deficiency of the new Fisher rule that also applies to the old one: the rule may mischaracterize pro-competitive behavior as predatory and anti-competitive. Although some anti-competitive actions have the features described by the Fisher rule, so do many pro-competitive actions. That is, every action that obeys the Fisher rule is not anti-competitive. There are many pro-competitive actions that firms take where they sacrifice profits for some time in order to gain profits in the long run. For example, suppose that, in an industry with network effects, a firm sells a product at a low price (but above incremental as well as break-even average cost), expecting to attract customers to its product of proprietary design. Moreover, suppose that as new customers come to market, the firm continues to use this strategy and does not increase the price over an extended period of time. As a consequence, the firm's market share increases to the point that this firm is considered dominant. This strategy may have purely pro-competitive causes but it fits well in the Fisher definition of anti-competitive acts. A court implementing the Fisher rule could find such a firm liable of predation and thus eliminate the benefits to consumers of the pro-competitive strategy.

Finally, the Fisher rule does not require that the lost monopoly rents be eventually recovered. The courts are urged to characterize acts that fit the Fisher rule as predatory and to take action against firms exhibiting such behavior typically *before* the final act of recovery of profits occurs, and therefore without knowledge or certainty that profit recovery will occur. Thus, the Fisher rule does not wait for the final act (recovery of lost rents) that could distinguish between pro-competitive and anti-competitive acts. It rushes to judgment before a certainty of an anti-competitive act. Professor Fisher is concerned that unless antitrust authorities act *before* consumers are harmed, "corrective action might come too late." But, if we follow Professor Fisher's rule, we can rely only on faith in the correctness and applicability of an oligopoly model to reach the conclusion that consumers will be harmed at some time in the future. There will be no certainty. And, in most hypothetical situations, given the multitude of oligopoly models with contradictory results, there cannot even be a consensus of a high probability that consumers will be hurt in the future. Thus, the Fisher rule will just create one more battle ground among economic consultants, some stating that it is inevitable that consumers will be hurt in the future, while others seeing only pro-competitive behavior. Ultimately, if this rule is adopted, the big winners will be economic consultants. The big losers will be companies that take aggressive competitive actions, which will regularly be sued on predatory grounds by their less aggressive (or less efficient) rivals. In the long run, if the rule is widely adopted, repeated antitrust intervention will coach aggressive competitors not to be aggressive and to instead accommodate competitors, resulting in very significant losses to consumers.

### 9. On Tying Of Windows With Internet Explorer

As we have seen, the district court ruled that Microsoft had monopoly power in the operating systems market. Judge Jackson further ruled that the Internet browser market was separate from the market for operating systems, and that technological and contractual tying of Internet Explorer with Windows raised prices and hurt consumers. The judge ruled that consumers do not want IE even free of charge because it burdens the OS with memory and overhead requirements, and it consumes a significant amount of hard disk space.

There is a major contradiction in the district court's decision. Under the § 2 analysis, browsers are in the same market as operating systems, but in the Sherman §1 part of the case, browsers and operating systems are in different markets. On the one hand, in the tying part of the case, the judge ruled that Internet browsers are in a separate market from PC operating systems. On the other hand, in the monopolization part of the case, the judge ruled that Microsoft attempted to marginalize the Netscape browser to defend its Windows monopoly because the Netscape browser was (or was becoming) a close substitute for Windows. If Internet browsers are in a separate market from the operating systems market, then Microsoft's actions in the browser market in the monopolization part of the case cannot be interpreted as defending Microsoft's monopoly. If instead the browser is in the same market as Windows, there cannot be illegal tying.

Microsoft claimed that IE was giving crucial functionality to its operating system and had every right to include it in Windows under the terms of the 1995 consent decree, which allowed addition of functions and features to Windows. Although this is a legal matter, I believe that the consent decree's language is plain enough to allow one to conclude that Microsoft's view is correct.

But why did Microsoft produce a browser and why did it then give it away? Essentially the question is: what is the appropriate behavior in a complementary goods market for a firm that has a dominant position in one market? That is, does it make sense from a competitive point of view that Microsoft was giving away the browser at no charge, or was it an anti-competitive act? To understand this, we first outline some thoughts about extracting value from complementary goods chains, where one good has close complements that are used in conjunction with it.[39]

First, a firm in one market is better off and can extract more value when the complementary markets in which it does not participate are more competitive. Conversely, the more monopolized a market is, the less value remains for the complementary markets. Second, keep in mind that value can be extracted only *once* from the chain of components. If a firm can monopolize one component, and all other complementary component markets are perfectly competitive, the firm gains nothing by attempting to monopolize the complementary component markets. The value of a good or service cannot be extracted more than once, no matter how many components the good is broken into. But, a firm that monopolizes a component has an incentive to enter and compete hard in a market for a complementary component if the complementary component market is *not* perfectly competitive, because by entering such a market, the firm can capture rents that it was losing from its original market. Third, more generally, a firm that participates in a market that is *not* perfectly competitive has a stronger incentive to enter a (not perfectly competitive) market for a complementary component than a firm that does not participate in any other market. Otherwise put, when complementary component markets are not perfectly competitive, there are strong incentives for the same firms to enter more than one of these markets. Fourth, when a component, say component A, is used together with many other complementary components, say $B_1, ..., B_n$, to produce $n$ composite goods of varying market values, competitors in the market for A will tend to enter the markets for the complementary components that, in combination with A, have the highest market value.

---

[39] *See* Nicholas Economides, *Competition and Vertical Integration in the Computing Industry*, in Competition, Innovation, and the Microsoft Monopoly: Antitrust in the Digital Marketplace, Jeffrey A. Eisenach & Thomas M. Lenard (eds.), Kluwer Acad. Publishers 1999, (visited Apr.23, 2001), at http://www.stern.nyu.edu/networks/98-11.pdf.

The entry of Microsoft in the browser market can be seen in the context of competition and multi-market participation in markets for complementary components. As long as the market for browsers was competitive and was shared by a number of browsers, and as long as the Internet was a small academic market, Microsoft had no significant business interest in it. When the browser market became dominated by Netscape, and, simultaneously, the Internet market appeared to be significantly larger, Microsoft integrated browser features in Windows and significantly improved its browser. Again, there were two reasons for this. First, Netscape had a dominant position in the browser market, thereby taking away from Microsoft's operating system profits to the extent that Windows was used together with the Navigator. Second, as the markets for Internet applications and electronic commerce exploded, the potential loss to Microsoft from not having a top browser increased significantly. These reasons were sufficient to drive Microsoft to produce a browser.

Was the free distribution of IE anti-competitive? Clearly, Microsoft had a pro-competitive incentive to freely distribute IE since that would stimulate demand for the Windows platform. Were there anti-competitive effects? If one assumes (as DOJ does in the tying part of the case) that the two markets (operating systems and browsers) are separate and therefore tying occurred, the crucial question is whether tying was anti-competitive. In particular, was the incremental increase in the price of Windows when IE was bundled with it larger than the price increase justified by the value of functionality that IE added? This is the test that the Jackson court should have conducted, and it failed to do so. My view is that it is very hard to prove that a quality-adjusted new version of Windows *without* IE would not have had a higher hedonic price than old Windows. That is, a modest increase in the market price of Windows was likely to be justified by the enhancement of features of Windows even without the inclusion of IE. Thus, if the court had performed this test, I believe it is likely that it would have found that adding IE functionality to new versions of Windows and distributing IE free of charge for older versions of Windows and for other operating systems did not harm consumers.

**10. On Attempting To Monopolize The Browser Market**

The district court judge found that Microsoft lost money while trying to gain market share from Netscape, in its attempt to monopolize the browser market by leveraging its monopoly of the OS market. The judge also found that Microsoft failed in this attempt.

It is well understood that the potential surplus of a consumer for a unit of a good can only be extracted once. If Microsoft is able to extract the surplus of a consumer that buys an Internet browser through Windows, it has no incentive to try to monopolize the market for browsers, as long as the market for browsers is competitive. To the extent that the market for browsers is monopolized, it can take away from the surplus appropriated by Windows, and then Microsoft may have an incentive to expand its market share in browsers. There is significant evidence that the market for browsers was originally monopolized by Netscape. Then the inroad of Microsoft into browsers replaced a monopoly with duopoly. And, it brought the price to zero and intensified research and development in browsers to the great benefit of consumers. It is quite ironic that, although the inroad of Microsoft into browsers was greatly beneficial to consumers and broke the Netscape monopoly, the government and the judge found Microsoft's actions anti-competitive.

### 11. Other Potentially Anti-competitive Arrangements

The court proceeding spent a significant amount of time on potentially anti-competitive actions of Microsoft in its dealings with buyers, content providers and others. An examination of the various arrangements and the facts of each one of them is impossible in this article due to of lack of space. Broadly speaking, the potentially anti-competitive actions were of three kinds:

1. Contracts with OEMs on the distribution of Netscape and IE, and the lack of control by OEMs of the opening screen ("first screen") of a new PC.
2. Contracts with ISPs (such as AOL) on the distribution of Netscape and IE
3. Control by Microsoft or OEMs of "active channels" to be provided to Internet Content Providers (such as Disney)

The potentially anti-competitive effects of such contracts have to be judged on a case by case basis.

### 12. Effects On Consumers

In principle, there are three ways that consumers could be hurt by potentially anti-competitive actions. First, consumers may be hurt because these actions increase prices. Second, consumers may be hurt because these actions may limit their choices in terms of variety and quality. Third, these actions may limit innovative activity, thereby hurting future consumers. Any harm in any of these three dimensions should be evaluated and balanced with any benefit of these actions in other dimensions. It is customary in antitrust cases to value the benefits to consumers and balance them with the burden of anti-competitive actions. The district court failed to do this task. The court did not attempt to quantify the benefits to consumers from giving away Internet Explorer and the intensification of competition that resulted. Also, the court did not try to assign a monetary value to the losses to consumers resulting from the anti-competitive actions for which it found Microsoft liable.

Consumers have directly benefited from the free distribution of Internet Explorer as well as its bundling and tight integration with Windows. Before Microsoft started to seriously compete with Netscape in the Internet browser market, Netscape— essentially the sole provider of Internet browser software—charged non-academic users $40-50 to use its browser.[40] Microsoft, by contrast, gave its Internet browser away. Netscape responded to the introduction of IE version 4 that many held of superior or comparable quality to Netscape by giving its browser away as well. Today, with at least 100 million browsers installed in the United States, Microsoft's actions have created a benefit of at least $4 to $5 billion to American consumers. And, since Microsoft's actions intensified competition, which in turn produced higher quality browsers, they provided further benefits to consumers.

The district court judge ruled that the prices of Windows 95 and 98 were too high, based on internal Microsoft e-mails that discussed a range of prices considered by the company for pricing Windows. However, consumers may have directly benefited from the relatively low price of Windows. Since the marginal cost of Windows is almost zero, clearly Microsoft had a wide range of prices it could have used. The static monopoly price for Windows was thirty to forty times higher. Microsoft's operating system, for which computer manufacturers pay $40-60 per copy, is cheap compared to the historical and current prices of other operating systems. For example, in the late 1980s, IBM sold OS/2 (which ran much fewer applications than Windows) for hundreds of dollars. Some Linux packages—essentially add-ons to the free Linux source code – currently sell for up to $150, and run far fewer applications than Windows does. These price discrepancies highlight a huge contradiction in the government's case and in the judge's findings of fact. If Microsoft were a true malevolent monopoly unconstrained by potential competition, it would charge far more for Windows than it does. The annual consumer benefits from Windows' relatively low price may be many billions of dollars.

The district court judge ruled that Microsoft's action of distributing Internet Explorer at no charge "increased [the] general familiarity with the Internet and reduced the cost to the public of gaining access to it," "gave Netscape an incentive to improve Navigator's quality," and benefited consumers since it "compelled Netscape to stop charging for Navigator."[41] On the other hand, the district court judge ruled that consumers were hurt in various ways by Microsoft's anti-competitive actions. The judge ruled that consumers were hurt[42] because they could not get Windows without a browser since Microsoft "forced OEMs to ignore consumer demand for a browserless version of Windows," and that the inclusion of Internet Explorer led to "degraded system performance, and restricted memory."

---

[40] Netscape was distributed free of charge to academic and governmental institutions. Netscape was also distributed free of charge to all in its introductory stage.

[41] Findings of Fact, United States v. Microsoft, Civil Action No. 98-1232 (TPJ) and State of N.Y. v. Microsoft Civil Action No. 98-1233 (TPJ), thereafter ("FOF"). *See* Microsoft III, 84 F. Supp 2d at 111. (paragraph 408 discussing the effect on consumers of Microsoft efforts to protect the applications barrier to entry).

I believe that Judge Jackson erred in ruling that free distribution of IE significantly harmed consumers by degrading performance and restricting available computer memory. First, because IE has been integrated into Windows, it is very hard to judge how a computer would work without it, how much memory it would use, how quickly it would run, etc. One would have to construct a computer with some other new component of the OS taking over the functions that IE performs now in Windows, and then test it. The court failed to do so, and cannot conclusively rule on this issue. Second, the Internet-related tasks of operating systems are in very high demand as the Internet is presently in an exponential expansion path. The number of consumers who have been harmed in the way suggested by the judge is likely to be very small, and in an age of cheap computers, memory, and hard drive capacity, their losses could not be very large.

Netscape and Internet Explorer ended up being very similar products in terms of their functionality. Most computer magazines rate recent versions of IE above Netscape's. There does not seem to be any loss in variety or quality by the domination of the browser market by IE. Windows-based computers can run both IE and Navigator simultaneously, and users are not forced to choose one to the exclusion of the other. One could use one browser for some tasks and another for others.

The court ignored the issue of compatibility which was probably the most central issue in the case. Backward and forward compatibility is crucial for software markets, and, as explained earlier, is the source of network effects. Microsoft provides *de facto* compatibility through its Windows operating systems. Compatibility is an important benefit to society because it is the source of network effects. Compatibility and its benefits could be quickly eliminated or significantly reduced if Microsoft is broken into competing pieces. Microsoft did not forcefully present to the court the benefits to society of its dominant position arising from the de facto compatibility it created and the resulting network effects.

Finally on the issue of innovation, economists' opinions are split on whether monopoly or competition would create more innovation. Economists' opinions are also split on whether vertically integrated or independent companies create more innovation. The government's economic witnesses did not provide any specific theory to support the claim that Microsoft was damaging the innovation process. The theory that general fear of Microsoft kept competitors out of the market or without available capital is contradicted by the abundant support by financial markets of the successful entry of a number of Linux variants.

---

[42] *See* Microsoft III, 84 F. Supp 2d at 111. (Findings of Fact discussion at paragraph 410).

**13. Remedies**
   **a.	Remedies As Imposed By The District Court**

The issue of remedies was from the very beginning the most difficult part of the litigation. Many observers that believed that Microsoft was liable had expressed privately great consternation in finding an appropriate remedy. Given the variety of remarks of Judge Jackson during the trial, there was no doubt that the district court would have found Microsoft liable. Even before the trial started, it seemed that the remedies determination would be the most crucial and most interesting element of the process. It was widely expected that the remedies phase would be a new full scale trial with significant input by economists.

Instead, Judge Jackson decided on remedies in summary fashion without any examination of the proposals of all sides. He was quoted shortly afterwards arguing that since the plaintiffs won the liability phase, the plaintiffs should have the right to determine the remedies.

In the last week of March 2000, the *New York Times* published the terms of the Microsoft and DOJ negotiations conducted under Judge Posner's supervision. The published information implied that DOJ and Microsoft were not far from an agreement, but that some States that were part of the litigation disagreed with the settlement terms. The terms of the final draft of the proposal of DOJ were:[43]

1.	Microsoft would create a pricing schedule that would apply to all buyers, so that prices would not be conditioned on other Microsoft products that a buyer buys. The schedule would allow for different prices for different quantities.
2.	Microsoft would not be allowed to have exclusive contracts that do not allow the other party to use, display, or feature its opponent's products.
3.	Microsoft would be required to share technical information without discrimination among the potential recipients of this information.
4.	Microsoft would be required to disclose the application interfaces (APIs) that link applications to Windows.
5.	Microsoft cannot increase the price of old versions of Windows when new versions are released. Microsoft would be forced to sell the old version at a constant price for three years after a new release.
6.	Microsoft would produce a Windows version without IE. Computer manufacturers would be allowed to license some part of the Windows code so that they could change the opening screen, and choose the default browser
7.	Tying by contract would be prohibited, but Microsoft would be allowed to integrate functions, applications, and features in its products.[44]

---

[43]	*See generally* N. Y. Times, Apr. 2, 2000. *See also* Brinkley & Lohr, *supra* note 7, at 286-87.

[44]	This seems identical to the provision of the 1995 decree which the DOJ claimed at the beginning of this lawsuit was not restricting Microsoft enough.

This list of conditions seemed a reasonable basis on which a deal with Microsoft could be struck. Since these were the government's proposals, one expects that the final negotiated settlement would have been somewhat less harsh on Microsoft. It was widely reported that some States officials considered these remedies as less harsh than they wished for. These States officials communicated to Judge Posner that they would not sign such a deal. Judge Posner was forced to declare the negotiations a failure.

Given the proposals that the government offered in the negotiations, there was a wide expectation that DOJ would demand more or less the same terms in the remedies phase.[45] Instead, DOJ asked for a much more radical step: a breakup of Microsoft.

The plaintiffs remedies proposal as adopted by the judge imposed a breakup of Microsoft into two "Baby Bills,"[46] an operating systems company which would inherit all the operating systems software, and an "applications" company with all the remaining software assets. Cash and securities holdings of other companies held by Microsoft would be split between the resulting entities. Bill Gates and other officers / shareholders of the company would not be allowed to hold executive and ownership positions in both of the resulting companies.

The district court ruling also imposed interim conduct restrictions on Microsoft. These restrictions, to last three years, from the time of the breakup were:

1. Microsoft would create a pricing schedule that would apply to all buyers, so that price would not be conditioned on the sale of other Microsoft products.
2. Microsoft would not be allowed to have exclusive contracts that do not allow the other party to use, display, or feature its opponent's products.
3. APIs and other technical information of Windows should be shared with outsiders as it is shared within Microsoft.
4. Microsoft is not allowed to take actions against manufacturers who feature competitors' software.
5. Microsoft will allow OEMs to alter Windows in significant ways.
6. Microsoft is not allowed to design Windows to disable or compromise rivals' products.

These conditions are similar but more restrictive than the ones proposed by the government in the settlement talks at the end of March 2000. At that point in time the government was satisfied that these restrictions, together with a ban on tying and a requirement that Microsoft would produce a version of Windows without IE, were sufficient to settle the case. DOJ did not produce any significant reason that made it change its position. Moreover, DOJ did not produce any convincing argument that the proposed breakup would be more efficient or effective than the conduct restrictions that it had proposed earlier.

---

[45] In fact, Professor Harry First who was acting as Assistant Attorney General for Antitrust for the State of New York was quoted in the New York Times (Apr. 30, 2000) saying that he and other officials of the States were expecting DOJ to propose remedies similar to DOJ's proposals at the negotiations, and were very surprised when Assistant Attorney General Joel Klein informed them a few days before the filing that DOJ would propose a breakup. *See also* Brinkley & Lohr, supra note 7, at 306-07.

[46] This is a word play on "Baby Bells" that came out of AT&T and the first name of the CEO of Microsoft, Bill Gates.

In arguing for the break-up, the government put forward a number of reasons. But, since there was only a summary formal hearing on remedies, there was no chance for both the government's and Microsoft's cases on remedies to be discussed and evaluated. The government and the judge have stated (formally and informally) the following arguments for a breakup:

1. That it considered the repeated violations of antitrust law by Microsoft as an indication that Microsoft would not follow any conduct or contractual restrictions; in fact, in some informal remarks, government officials believe that they were "tricked" by Microsoft in settling the 1995 case with terms that Microsoft was able to exploit;
2. That the lack of remorse by Microsoft's executives was a clear indication that Microsoft "could not be trusted" to implement any other remedy;
3. That the breakup was a "surgical cut" that would create the least interference with business;
4. AT&T and the rest of the telecommunications industry benefited from AT&T's breakup, and so should Microsoft and the software industry; after all both industries have network effects;
5. The breakup eliminates the incentive for vertical foreclosure; and
6. The breakup reduces the "applications barrier to entry" since now the applications company might write popular Microsoft applications (such as MS-Office) for other platforms.

I believe that the government has failed to show that the proposed breakup is the appropriate remedy. DOJ had to show that conduct remedies would not work and it failed to do so. DOJ has not performed the appropriate cost-benefit analysis to show that conduct remedies are not sufficient and that a breakup is necessary. None of the affidavits in the remedies phase even approach a discussion on evaluating alternatives. As discussed earlier, the government also failed to justify why it was ready to compromise a few weeks earlier on behavioral remedies but now it claimed that structural remedies were necessary.

The first argument of the government does not stand to reason. The 1995 case was settled with a decree that explicitly stated that Microsoft can include in its operating system *any* additional functionality. It is reasonable that Microsoft (or any observer, including DOJ) would believe, given the 1995 consent decree, that adding browser functionality to Windows does not violate the consent decree. This, of course, does not mean that adding such functionality does not violate antitrust law in general, but it puts to its death the idea that the government was tricked by Microsoft. Of course, it stands to reason that Microsoft understood much more clearly what the degree said than DOJ since Microsoft understood the business better and could predict what functions could be included in the operating system. However, the fact that companies and antitrust enforcers often have an asymmetry of information is very common and expected, and cannot be considered a "trick," or a reason not to enter into agreements between antitrust authorities and companies.

The second argument of the plaintiffs makes no sense at all. Antitrust enforcement is not an emotional tug of war in which the egos of either the plaintiffs or the defendants need to be satisfied. The show of remorse or lack thereof by Microsoft executives could not possibly define the remedy. I find it hard to believe that the judge would really find a different remedy appropriate if enough Microsoft executives simply showed public remorse. Moreover, Microsoft, like any other defendant, has a right to appeal. Belief that it will prevail on appeal in a civil case is hardly worthy of punishment.

The third argument, that the breakup is a surgical cut and therefore will disrupt the industry the least is countered by the facts. A breakup of Microsoft, if it were finally ordered at the end of the appeals process, would, practically speaking, eliminate Microsoft as a flexible and formidable competitor. The wholehearted endorsement of the breakup by Microsoft's competitors in the servers and backoffice competitors (who were not alleged to have been damaged by the Windows monopoly but will greatly benefit from the confusion and disruption created by a Microsoft breakup) is evidence that the breakup is one of the most disruptive possible outcomes.

The fourth argument, that since AT&T's 1982 breakup was successful so would Microsoft's, is incorrect. AT&T was divided into the long-distance company (AT&T), and seven regional operating companies, each of which remained a regulated local telecommunications monopoly until 1996. The destruction of AT&T's long-distance monopoly encouraged competition, which brought sharply lower prices and immense consumer benefits.[47] There are a number of key differences between the two companies and their competitive situations. And these differences make it very likely that a Microsoft breakup, besides harming Microsoft, would harm consumers and the computer industry.

In 1981, AT&T was a 100-year-old regulated monopoly with many layers of management. For historical reasons, the local phone companies within the old AT&T, such as New York Telephone, were managed separately from the "long lines" division. Thus, it was not difficult to separate the divisions since they functioned on many levels as separate companies. AT&T also had an abundance of managers to help cope with the breakup. By contrast, Microsoft is a young, entrepreneurial company run by very few top executives (about 25), and its divisions are very fluid. While this has made Microsoft one of the most efficient and successful companies around, it also means that a break-up would pose significant managerial problems and severely reduce the company's flexibility. Finally, AT&T was a regulated utility and regulation guaranteed that the companies emerging from the breakup stayed interconnected. In contrast, the Microsoft breakup is likely to lead to incompatibilities and further loss of efficiency.

---

[47] At the same time the AT&T breakup did not introduce competition at the local exchange level, and the Regional Bell Operating Companies ("RBOCs") were allowed to monopolize local telecommunications services as well as access to long distance services. The success of competition in long distance has been hampered by the continuing monopoly of the local exchange, four years after the Telecommunications Act of 1996 was supposed to open the local exchange to competition.

DOJ's two-way breakup plan was premised on the hope that an autonomous applications company would create a new operating system to compete with Windows. But more than 70,000 applications run on Windows, creating what the government calls "the applications barrier to entry" in the operating-system market. However capable the new applications company, it still wouldn't be able to single-handedly create a successful rival operating system. Separately, even with a new applications company's support, Microsoft's biggest operating-system competitor, Linux, is unlikely to become a serious desktop threat to Windows.

In my opinion, the breakup as proposed by the government and imposed by the district court will have detrimental effects. First, the breakup is likely to result in higher prices. If DOJ is correct and Microsoft kept its OS prices low so that it can exercise its monopoly power in the adjacent browser market, the post-breakup Baby Bill that inherits the operating systems will have no incentive to keep the price low. The OS Baby Bill will no longer have the incentive to disadvantage any applications companies. Thus, if DOJ's theories are correct, the OS Baby Bill will now exercise the monopoly power it has and raise the price of the operating system to the detriment of consumers. If DOJ is correct and Microsoft has significant monopoly power because of the "applications barrier to entry," higher prices will be the direct result of the breakup. Second, as explained earlier, the breakup is likely to eliminate the efficiencies that make Microsoft a flexible and formidable competitor.

The breakup is likely to temporarily eliminate the incentive for interference from OSs to applications and vice versa. Of course, the same could have been accomplished by conduct restrictions without the cost and the disruption of the breakup. Moreover, the district court did not impose permanent restrictions on the post-breakup functions of the companies. The OS and the applications Baby Bills may enter into each other's business soon after the breakup. It is very likely that a few years after the breakup, one of the resulting companies will dominate both markets.

### b.    Other Remedies Proposals

A number of other remedies proposal have been discussed in the literature, but unfortunately were not discussed in the district court's summary remedies proceedings. If the liability conviction is upheld by the court of appeals or the U. S. Supreme Court, it is likely that the issue of remedies will be re-opened and a full hearing on remedies may take place.

In principle, remedies fall into two broad classes. First there are those that attack the business side of Microsoft by restricting its contracts or splitting the company along various lines of business. The second broad class attacks the control over the technology by either forcing Microsoft to release or license the source code, or to disclose proprietary information (APIs). The remedies can also be divided into those that affect conduct and those that affect structure. For example, a restriction on contracts is a conduct remedy, while a breakup is a structural remedy.

Provided that the liability conviction is upheld, four alternatives have been proposed. One remedy proposal is even more extreme than the one imposed by the judge. This extreme proposal would break up Microsoft into three identical Baby Bills, with each company acquiring the source code of all the programs that Microsoft currently sells, and one third of its employees.[48] This "horizontal remedy" is sometimes presented in combination with the "vertical remedy" imposed by the judge. In this "hybrid remedy," first Microsoft is broken into two or three companies according to the type of program produced, and then the operating systems company is broken into three parts creating four of five companies altogether.

Besides the loss of flexibility that any breakup would create, a horizontal or hybrid breakup would also produce significant incompatibilities with harmful effects to computer users, applications writers, and Microsoft shareholders. Baby Bills coming out of a horizontal or a hybrid breakup will have incentives to create incompatible versions of Windows for two reasons. First, Baby Bills will try to differentiate their operating systems to avoid strong competition, leading to small price-cost margins. This is true even in industries without network externalities and has been well established in the economics literature on product differentiation.[49] Second, Baby Bills will try to make their operating systems incompatible with each other in a race to become the dominant OS, since the dominant firm receives the lion's share of profits in a winner-takes-most world. This is established in the network economics literature.[50] Differentiating the operating systems by Baby Bills would inevitably reduce the range of software that would be compatible with each user's computer. As a consequence, consumers' surplus would decrease. The emerging incompatibilities would be a huge headache for both independent applications writers and corporate IT departments. Such incompatibilities would also hurt shareholders, since the combined value of the resulting Baby Bills will be smaller than that of the original Microsoft.

---

[48] *See* Litan, *supra* note 21.

[49] *See* Claude D'Aspremont, Jean Jaskold-Gabszewicz, & Jacques-Francois Thisse, *On Hotelling's Stability in Competition*, 47 Econometrica at 1145-50. *See also* Anver Shaked & John Sutton, *Relaxing Price Competition Through Product Differentiation*, 49 Rev. Econ. Stud. at 3-14. *See* Nicholas Economides, *The Principle of Minimum Differentiation Revisited*, 24 Eur. Econ. Rev. at 345-368.

[50] *See* Economides & Flyer, *supra* note 20, and Nicholas Economides, *Industry Fragmentation After A Microsoft Breakup* (2001) (on file with the author); Nicholas Economides, *The Microsoft Antitrust Case*, J. Indus., Competition & Trade: From Theory to Policy (forthcoming Aug. 2001); *see also* Nicholas Economides, *The Microsoft Antitrust Case: Rejoinder*, J. Indus., Competition & Trade: From Theory to Policy (forthcoming August 2001).

Another remedy proposal is auctioning the Windows source code. Given the fluctuating stock market value of Microsoft, Windows source code may be worth as much as $200 billion. No company can bid that much cash in an auction. (Practically speaking, only a handful of foreign governments could). This implies that the source code of Windows would be sold forcibly at a small fraction of its worth – and that would severely reduce the value of shareholders' equity. Auctioning the Windows code would not only effectively confiscate Microsoft's intellectual property, it would also seriously reduce the incentive for innovation not only for Microsoft but for all potential innovators. Moreover, source code evolves. Over time, different firms will add and alter the Windows code. Soon, incompatibilities will arise, with all the negative consequences of diminution of network effects described earlier.

A second remedy proposal would be to impose on Microsoft to disclose some of the so-called "APIs" (lines of software code that define interfaces between applications) that permit it to include Internet Explorer in the operating system. Microsoft routinely discloses APIs that hook applications to the operating system and allow for interoperability. Currently, it does not disclose APIs that tie together parts of the Windows operating system, which includes Internet Explorer. If Microsoft were to disclose the APIs that hook Internet Explorer to other parts of the operating system, Netscape (or any other browser) could have the same interoperability with Windows.[51]

A third remedy proposal—and in my mind, the best—would be to consider imposing various restrictions on the contracts that Microsoft can write with sellers of complementary goods and with competitors. This is a likely remedy that is easy to tailor according to the violation. This remedy—possibly combined with requiring Microsoft to disclose certain APIs—should be sufficient to guarantee that Microsoft will be precluded from taking future anti-competitive actions. At the same time, it would preserve the managerial and other benefits that have made Microsoft one of the most successful and most profitable companies ever.

Some have proposed regulation of the operating systems part of Microsoft as a way to guarantee that there would be no leverage of market power. There are very significant reasons to avoid regulation at this stage. First, regulation is inappropriate for an industry that evolves quickly. Regulation requires a stable product. For example, regulation of AT&T did not occur until 35 years had passed after the invention of the telephone, despite the fact that (i) telecommunications had much stronger network effects than software; and (ii) during its first 40 years, AT&T took much more aggressive actions than Microsoft, including refusal to deal and refusal to interconnect with competitors. Second, regulation could stagnate innovation. Third, in an industry with fast technical progress, regulation can be used by the regulated companies to keep prices high, as exemplified by telecommunications regulation.

---

[51] This is a disclosure of a few APIs that hook IE to Windows. It does not imply a disclosure of a large number of APIs, which would be almost equivalent to disclosure of the Windows source code.

**14. Concluding Remarks**

The most significant recent developments in this case were the filing of the appeal of Microsoft, the filing of the government's objections, and the open hearing of the case by the Court of Appeals for the D.C. Circuit sitting *en banc* in late February 2001. The questioning of both sides by the judges of the court of appeals was very frank and pointed. Most of the criticism of the appeals judges (as seen by their questions and by the heated dialog with the lawyers of both sides) was aimed towards the plaintiffs. On a number of occasions, the appeals court judges questioned the validity of the district court's "findings of fact," calling them conclusions not based on fact.[51] The appeals court also seemed to question the strength of the plaintiffs' case (i) on the tying allegation, as well as (ii) on attempting to monopolize the browser market allegation.

The appeals court was critical of the procedure that Judge Jackson used in the remedies part of the case,[52] as well as of the necessity and effectiveness of the ordered remedy (breakup).[53] The appeals court also seemed to be very critical of Judge Jackson's

---

[51] For example, at the February 27, 2001 morning hearing, judges questioned Jeffrey P. Minear, representing the United States, as follows:
> "THE COURT: The District Court, like I said, there are some findings that are merely just conclusions and I find no basis for them. So I'm not in that camp that says because the District Court lists something under findings of fact it's gospel. There has to be a fact in fact.
> THE COURT: It has to be supported.
> THE COURT: And it has to be supported by something other than the mere statement of the District Court."

In a second example from the same hearing, a judge asks Mr. Minear:
> "THE COURT: Let me ask you again so that you can help me. This is one of the cases, one of the places for me, where the failure of the findings of fact to point to any record, citations, makes it very, very difficult on appellate review because they are very conclusionary statements here that I tried to trace to determine whether there was any real data to support the observation that there was a market for browserless operating systems. It is certainly not intuitive given that all of the operating systems offer browsers that can be removed or deleted.
>
> But in making your argument that in all the other cases they can be removed and therefore Microsoft is forcing, you're ignoring Microsoft's counter-argument which is they don't integrate as deeply.
>
> But in any event, make that your second answer. Tell me if there is any data to back up … I quite frankly ... I hear my colleagues in the first part of this argument that we're supposed to defer to factual findings. But when I find factual findings that look very conclusionary and there is no citation to anything, I don't think my obligation as an appellate court is to defer to them. So what is the data?"

[52] For example, at the February 27, 2001 morning hearing, a judge asked David C. Frederick, representing the United States:
> "THE COURT: Let me ask you a couple of questions about the standard applied by the District Court. The District Court said that the plaintiffs won the case, and for that reason alone have some entitlement to a remedy of their choice. The District Court also said these officials are by reason of office obliged and expected to consider and to act in the public interest. Microsoft is not. Are those appropriate standards for a District Judge to consider in framing a remedy?"

[53] For example, at the February 27, 2001 morning hearing, a judge pointed to David C. Frederick, representing the United States:

extended interviews with journalists and strongly suggested that if the case is remanded to a lower court (for either a new determination of facts, or for determination of remedies, or both) that it should not go to Judge Jackson.[54] Overall, the court of appeals appeared unlikely to affirm all three parts of the liability decision of the lower court and also unlikely to affirm the breakup remedy.

It is almost certain that the appeals court will reverse the tying decision and will find Microsoft not liable on tying. In my opinion, it is likely that the major parts of the rest of the district court's decision would also be reversed. After review by the appeals court and possibly by the U. S. Supreme Court, it is extremely unlikely that the final outcome will be a breakup of Microsoft. In fact, the final remedy imposed on Microsoft or the terms of a possible settlement of the DOJ suit are very likely to be weaker than DOJ's settlement terms of March 30, 2000, to which Microsoft is reported to have agreed.[55]

It is most likely that the federal part of the case will be settled after the court of appeals decision. Microsoft will have very significant legal fees and related costs, as it also faces a large number of civil antitrust suits by class action plaintiffs. The breakup would be disastrous, but, in my opinion, is extremely unlikely to be upheld. The biggest loss to Microsoft is the continuous antitrust scrutiny that does not allow it to make significant acquisitions in telecommunications and the Internet in the United States during the period of intense antitrust scrutiny.

The antitrust trial has put a tremendous pressure on Microsoft. If the company was not controlled by Bill Gates and employees, it is very likely that, in the short-term interest of the shareholders, Microsoft would have compromised with DOJ, even though this is likely to have harmed the long-term interests of shareholders. As things stand in May 2001, the stock price of Microsoft is down about 40 percent because of the continuing antitrust uncertainty.

---

> "THE COURT: Stranger still, even after the remedy, Microsoft retains the monopoly. You cited Grinnell, and I think there's a point in the Supreme Court's opinion in Grinnell that says the first order of business when there's been a Section 2 violation is to issue a remedy that will destroy the monopoly power. This remedy doesn't do that."

In another question in the same hearing, a judge notes the potential problems that a breakup might create and the fact that there was no hearing to discuss them:

> "THE COURT: No, no, no. The question that's being raised is whether a company that has not grown through combinations can be perforated along the lines proposed by the government without a hearing into the problems that might create."

[54]   For example, at the February 27, 2001 afternoon hearing , a judge asked John G. Roberts, representing the States plaintiffs:

> "THE COURT: Well, I'm not sure that I see how you can with a straight face ask us if we remand, to send it to the same judge after these comments."

[55]   Provided of course that published reports of the settlement proposal in the New York Times and elsewhere are correct as summarized..

Regardless of the final outcome, the effects of U.S. v. Microsoft are likely to be felt for a long time. If a breakup actually occurs, it is likely to impose the dark shadow of radical antitrust intervention on the whole computer industry. There are many firms in the computer sector that have a dominant position in their respective markets. I would expect, if on appeal DOJ wins big on Microsoft (which is very unlikely) and if DOJ takes a principled approach, antitrust suits against Yahoo, AOL, and other pioneers of the New Economy will not be far behind.

**15. Bibliography**

Amicus Curiae Brief Of Professor Lawrence Lessig, United States v. Microsoft, Civil Action No. 98-1232 (TPJ).

Areeda, Phillip and Louis Kaplow (1997), *Antitrust Analysis*, Aspen Law & Business.

Brinkley, Joel and Steve Lohr (2000), *U.S. v.* Microsoft, McGraw Hill.

Cass, Ronald A., (1999), "Copyright, Licensing, And The 'First Screen'," mimeo., Boston University, School of Law.

Cass, Ronald A., and Keith N. Hylton, (1999), "Preserving Competition: Economics Analysis, Legal Standards And Microsoft," Boston University, School of Law, Law & Economics Working paper no. 99-1.

Conclusions Of Law, United States v. Microsoft, Civil Action No. 98-1232 (TPJ) & State of New York v. Microsoft Civil Action No. 98-1233 (TPJ).

Davis, Steven J., Jack MacCrisken, and Kevin M. Murphy, (1998), Integrating New Features Into The PC Operating System: Benefits, Timing, And Effects On Innovation," mimeo.

D'Aspremont, Claude, Jaskold-Gabszewicz, Jean, and Thisse, Jacques-Francois, (1979), "On Hotelling's 'Stability in Competition'," *Econometrica*, vol. 47, pp. 1145-1150.

Economides, Nicholas (1984), The Principle of Minimum Differentiation Revisited," *European Economic Review*, vol. 24, pp. 345-368.

Economides, Nicholas (1996), "The Economics of Networks," *International Journal of Industrial Organization* vol. 14, no. 2, pp. 675-699, at http://www.stern.nyu.edu/networks/top.html.

Economides, Nicholas (1999), "Competition and Vertical Integration in the Computing Industry," in *Competition, Innovation, and the Microsoft Monopoly: Antitrust in the Digital Marketplace*, Jeffrey A. Eisenach and Thomas M. Lenard (eds.), Kluwer Academic Publishers 1999, at http://www.stern.nyu.edu/networks/98-11.pdf. .

Economides, Nicholas (2001a), "Industry Fragmentation After A Microsoft Breakup," mimeo.

Economides, Nicholas (2001b), "The Microsoft Antitrust Case" *Journal of Industry, Competition and Trade: From Theory to Policy* (August 2001).

2001] ISSUES OF THE MICROSOFT ANTITRUST 43Economides, Nicholas (2001c), "The Microsoft Antitrust Case: Rejoinder" *Journal of Industry, Competition and Trade: From Theory to Policy* (August 2001).

Economides, Nicholas, and Fredrick Flyer (1998), "Compatibility and Market Structure for Network Goods," Discussion Paper EC-98-02, Stern School of Business, N.Y.U., http://www.stern.nyu.edu/networks/98-02.pdf .

Elzinga, Kenneth G., and David E. Mills (1999), "PC Software," *The Antitrust Bulletin*, vol. 44, page 739.

Evans, David S., and Richard L. Schmalensee, "Be Nice to Your Rivals: How the Government is Selling an Antitrust Case Without Consumer Harm in United States v. Microsoft," in *Did Microsoft Harm Consumers? Two Opposing Views*, American Enterprise Institute-Brookings Joint Center for Regulatory Studies, Washington DC 2000.

Evans, David S., Albert Nichols, and Bernard Reddy, (1999), "The Rise And Fall Of Leaders In Personal Computer Software," mimeo., NERA.

Findings of Fact, United States v. Microsoft, Civil Action No. 98-1232 (TPJ) & State of New York v. Microsoft Civil Action No. 98-1233 (TPJ).

Fisher, Franklin M., Written testimony, Civil Action No. 98-1232 (TPJ) & State of New York v. Microsoft Civil Action No. 98-1233 (TPJ).

Fisher, Franklin M. (2001), "Innovative Industries And Antitrust: Implications Of The *Microsoft* Case," *Journal of Industry, Competition and Trade: From Theory to Policy* (August 2001).

Fisher, Franklin M., and Daniel L. Rubinfeld, "United States v. Microsoft: An Economic Analysis," in *Did Microsoft Harm Consumers? Two Opposing Views,* American Enterprise Institute-Brookings Joint Center for Regulatory Studies, Washington DC 2000.

Hatch, Orrin G. (1999), "Antitrust in the Digital Age," in *Competition, Innovation, and the Microsoft Monopoly: Antitrust in the Digital Marketplace*, Jeffrey A. Eisenach and Thomas M. Lenard (eds.), Kluwer Academic Publishers 1999.

Klein, Benjamin, (1999), "Microsoft's Use Of Zero Price Bundling To Fight The 'Browser Wars'," in *Competition, Innovation, and the Microsoft Monopoly: Antitrust in the Digital Marketplace*, Jeffrey A. Eisenach and Thomas M. Lenard (eds.), Kluwer Academic Publishers 1999.

Lenard, Thomas M., (2000), "Creating Competition in the Market for Operating Systems: A Structural Remedy for Microsoft," mimeo. The Progress & Freedom Foundation.

Levinson, Robert J., R. Craig Romaine, and Steven C. Salop, (1999), "The Flawed Fragmentation Critique of Structural Remedies in the Microsoft Case," Georgetown University Law Center, Working Paper No. 204874.

Litan, Robert E., Roger G. Noll, William D. Nordhaus, and Frederic Scherer, "Remedies Brief Of Amici Curiae In Civil Action No. 98-1232 (TPJ)."

McKenzie, Richard B. (2000), *Trust on Trial*, Perseus Publishing.

Reback, Gary L. *et al.*, Memorandum Of Amici Curiae In Opposition To Proposed Final Judgment, Civil Action No. 94-1564.

Plaintiffs' Memorandum In Support Of Proposed Final Judgment, United States v. Microsoft, Civil Action No. 98-1232 (TPJ).

Reddy, Bernard, David Evans and Albert Nichols (2000), "Why Does Microsoft Charge So Little For Windows?," mimeo.

Romaine, R. Craig and Steven C. Salop (1999), "Alternative Remedies for Monopolization in the Microsoft Case," mimeo.

Shaked, Avner and John Sutton, (1982), "Relaxing Price Competition Through Product Differentiation," *Review of Economic Studies*, vol. 49, pp. 3-14.

United States v. Microsoft, conference at New York University (May 5, 2000), in streaming video at http://www.stern.nyu.edu/networks/video.html.

United States Court Of Appeals for the D. C. Circuit (2001), transcript of oral argument in United States v. Microsoft, case no. 00.5213, February 26-27, 2001.